\documentclass[aps,prd,twocolumn,showpacs,amsmath,amssymb]{revtex4-1}
\usepackage{amsmath}
\usepackage{graphicx}
\usepackage{subfigure}
\usepackage{epstopdf}
\usepackage{color}
\usepackage{multirow}
\usepackage{setspace}
\usepackage{overpic}
\usepackage{amssymb}
\usepackage[bookmarksnumbered, pdfstartview=FitH,colorlinks,urlcolor=blue, citecolor=blue,linkcolor=blue] {hyperref}
\usepackage{lineno}
\usepackage{bm}
\usepackage{rotating}
\usepackage[utf8]{inputenc}
\usepackage{amsmath}

\let\oldequation\equation
\let\oldendequation\endequation

\renewenvironment{equation}
  {\linenomathNonumbers\oldequation}
  {\oldendequation\endlinenomath}

\newcommand{\dstoKev}{D_s^+\to K_{1}(1270)^{0} e^+\nu_{e}}
\newcommand{\dstobev}{D_s^+\to b_{1}(1235)^{0} e^+\nu_{e}}
\newcommand{\koz}{K_{1}(1270)^{0}}

\begin{document}

\title{\bf \boldmath
Search for the semileptonic decays $D^+_s \to K_1(1270)^0 e^+\nu_e$ and  $D^+_s \to b_1(1235)^0 e^+\nu_e$}

%% Saved at => 2023-03-01
\author{
M.~Ablikim$^{1}$, M.~N.~Achasov$^{5,b}$, P.~Adlarson$^{75}$, X.~C.~Ai$^{81}$, R.~Aliberti$^{36}$, A.~Amoroso$^{74A,74C}$, M.~R.~An$^{40}$, Q.~An$^{71,58}$, Y.~Bai$^{57}$, O.~Bakina$^{37}$
% ...
\\
\vspace{0.2cm}
(BESIII Collaboration)\\
\vspace{0.2cm} {\it
$^{1}$ Institute of High Energy Physics, Beijing 100049, People's Republic of China\\
% ...
$^{5}$ Budker Institute of Nuclear Physics SB RAS (BINP), Novosibirsk 630090, Russia\\
% ...
$^{36}$ Johannes Gutenberg University of Mainz, Johann-Joachim-Becher-Weg 45, D-55099 Mainz, Germany\\
$^{37}$ Joint Institute for Nuclear Research, 141980 Dubna, Moscow region, Russia\\
% ...
$^{40}$ Liaoning Normal University, Dalian 116029, People's Republic of China\\
% ...
$^{57}$ Southeast University, Nanjing 211100, People's Republic of China\\
$^{58}$ State Key Laboratory of Particle Detection and Electronics, Beijing 100049, Hefei 230026, People's Republic of China\\
% ...
$^{71}$ University of Science and Technology of China, Hefei 230026, People's Republic of China\\
% ...
$^{74}$ University of Turin and INFN, (A)University of Turin, I-10125, Turin, Italy; (B)University of Eastern Piedmont, I-15121, Alessandria, Italy; (C)INFN, I-10125, Turin, Italy\\
$^{75}$ Uppsala University, Box 516, SE-75120 Uppsala, Sweden\\
% ...
$^{81}$ Zhengzhou University, Zhengzhou 450001, People's Republic of China\\
\vspace{0.2cm}
% ...
$^{b}$ Also at the Novosibirsk State University, Novosibirsk, 630090, Russia\\
% ...
}
}
%% ends here %%

\begin{abstract}
By analyzing 7.33\,fb$^{-1}$  of $e^+e^-$ collision data
collected at center-of-mass energies between 4.128 and 4.226 GeV with the BESIII detector,
we search for the semileptonic decays $D^+_s \to K_1(1270)^0 e^+\nu_e$ and  $D^+_s \to b_1(1235)^0 e^+\nu_e$ for the first time.
No significant signals are observed for either decay mode.
The upper limits on the (product) branching fractions are determined to be
${\mathcal B}[D^+_s \to K_1(1270)^0 e^+\nu_e] < 4.1\times
10^{-4}$ and ${\mathcal B}[D^+_s \to b_1(1235)^0 e^+\nu_e]\cdot {\mathcal B}[b_1(1235)^0\to \omega\pi^0] < 6.4\times 10^{-4}$ at 90\% confidence level.
\end{abstract}

\maketitle

\oddsidemargin  -0.2cm
\evensidemargin -0.2cm

\section{Introduction}

In the standard model (SM), semileptonic charmed meson decays provide an outstanding
probe to explore the dynamics of both weak and strong
interactions in the charm sector\cite{Li:2021iwf}.
Compared to the semileptonic decays into pseudoscalar and vector mesons,
charmed meson decays involving axial-vector mesons in the final state are not well-studied from either the experimental~\cite{pdg2020} or the theoretical~\cite{isgw,isgw2,khosravi,zuo,cheng,momeni1} side. Extensive studies of the semileptonic charmed meson decays into axial-vector
mesons $K_1(1270)$ and $b_1(1235)$ play an important role in
the understanding of non-perturbative strong-interaction dynamics in weak decays~\cite{isgw,isgw2,khosravi,zuo,cheng,momeni1}.
BESIII and CLEO collaborations have reported studies of $D^{0(+)}\to \bar {K}_{1}(1270) e^+\nu_{e}$~\cite{bes3-Dp-K1ev,bes3-D0-K1ev,cleo-D0-K1ev}. The reported branching fractions are consistent with theoretical predictions based on the Isgur-Scora-Grinstein-Wise (ISGW) quark model~\cite{isgw} and its updated version (ISGW2)~\cite{isgw2}, as well as with those based on the covariant light-front quark model~\cite{cheng}. The BESIII collaboration has also performed a search for $D^{0(+)}\to b_{1}(1235) e^+\nu_{e}$, and no significant signal was observed~\cite{bes3_D_b1ev}. For semileptonic $D_s^+$ decays into axial-vector mesons, no experimental study has been carried out so far.
\begin{figure}[htbp]
\centering
\includegraphics[height=3.7cm,width=6.55cm]{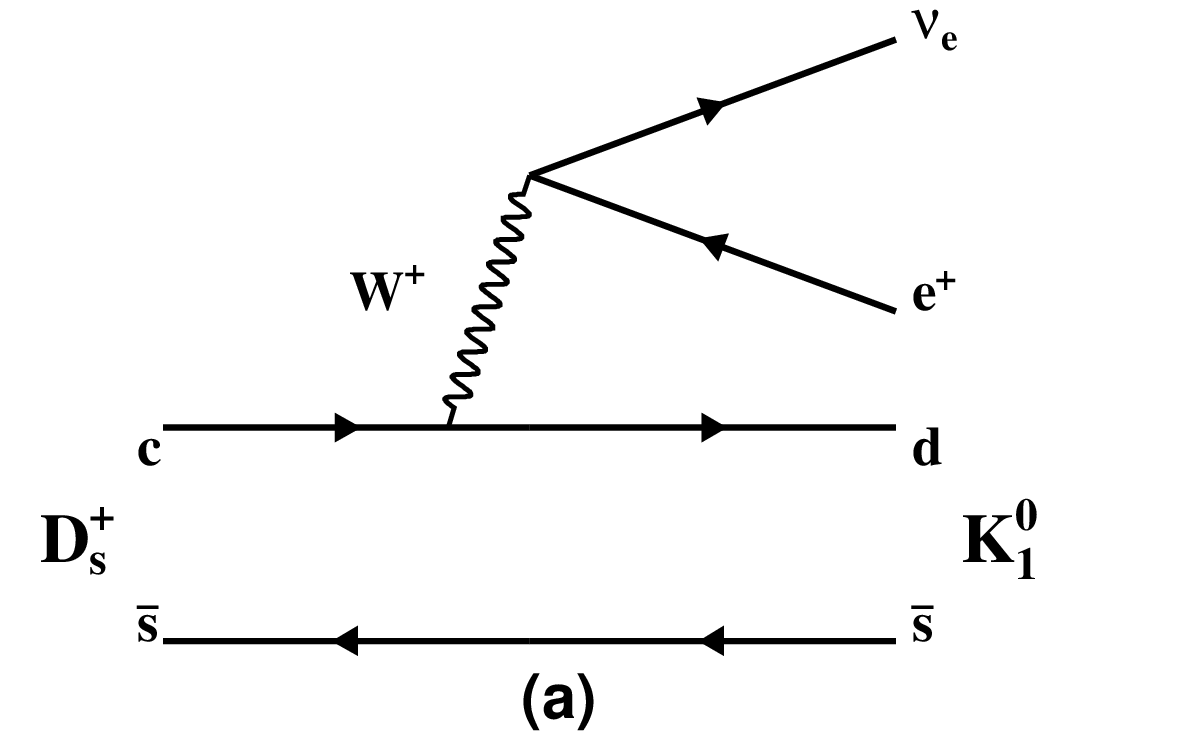}\\
\vspace{0.4cm}
\includegraphics[height=3.7cm,width=6.55cm]{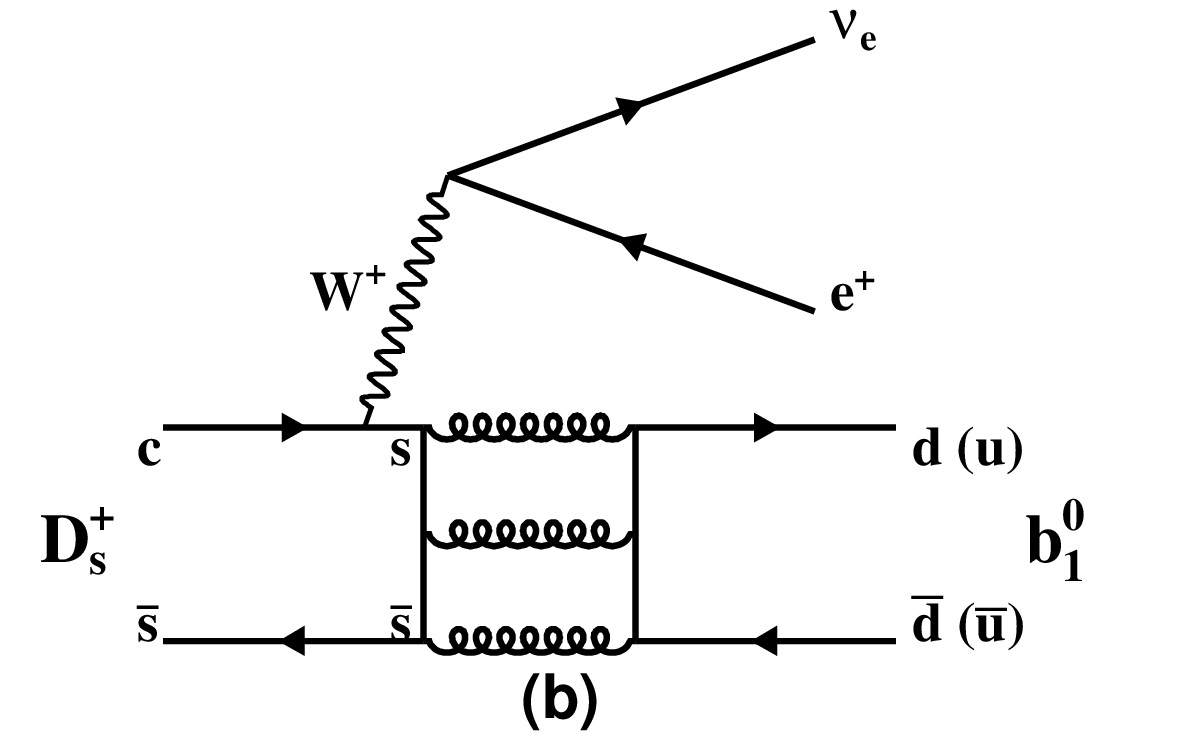}
\caption{Feynman diagrams of (a) $D^+_s \to K_1(1270)^0 e^+\nu_e$ and (b) $D^+_s \to b_1(1235)^0 e^+\nu_e$.}
\label{fig:fey}
\end{figure}

The semileptonic decays of $D^+_s \to K_1(1270)^0 e^+\nu_e$ and $D^+_s \to b_1(1235)^0 e^+\nu_e$ can proceed via the Feynman diagrams shown in Fig.~\ref{fig:fey}. Reference~\cite{cheng} has predicted the branching fraction of $D^+_s \to K_1(1270)^0 e^+\nu_e$ to be of order $10^{-4}$.
In contrast, the $D^+_s \to b_1(1235)^0 e^+\nu_e$ decay rate is highly suppressed due to isospin violation and the Okubo-Zweig-Iizuka (OZI) rule and is expected to be smaller than that of $\dstoKev$.
Experimental measurements of
% $D^+_s \to K_1(1270)^0 e^+\nu_e$ and  $D^+_s \to b_1(1235)^0 e^+\nu_e$
these two decay modes
are important to test theoretical calculations and to understand non-perturbative effects in heavy meson decays.

In this article, we present the first search for $D^+_s \to K_1(1270)^0 e^+\nu_e$ and  $D^+_s \to b_1(1235)^0 e^+\nu_e$ with $K_1(1270)^0\to K^-\pi^+\pi^0$ and $b_1(1235)^{0}\to\omega\pi^0$, respectively.
The charge conjugate channels are always implied throughout this paper.
This analysis is performed by analyzing $e^+e^-$ collision data corresponding to an integrated luminosity of 7.33\,fb$^{-1}$~\cite{lum}
collected at eight center-of-mass energy ($E_{\rm cm}$) points, as listed in Table~\ref{tab:mbc}, between 4.128 and 4.226 GeV with the BESIII detector.

\section{BESIII detector and Monte Carlo simulation}
The BESIII detector~\cite{Ablikim:2009aa} records symmetric $e^+e^-$ collisions
provided by the BEPCII storage ring~\cite{Yu:IPAC2016-TUYA01}
in the center-of-mass energy range from 2.0 to {4.95~GeV},
with a peak luminosity of $1 \times 10^{33}\;\text{cm}^{-2}\text{s}^{-1}$
achieved at $\sqrt{s} = 3.77\;\text{GeV}$.
BESIII has collected large data samples in this energy region~\cite{Ablikim:2019hff}.
The cylindrical core of the BESIII detector covers 93\% of the full solid angle and consists of a helium-based
 multilayer drift chamber~(MDC), a plastic scintillator time-of-flight
system~(TOF), and a CsI(Tl) electromagnetic calorimeter~(EMC),
which are all enclosed in a superconducting solenoidal magnet
providing a 1.0~T magnetic field. The solenoid is supported by an
octagonal flux-return yoke with resistive plate counter muon
identification modules interleaved with steel.
The charged-particle momentum resolution at $1~{\rm GeV}/c$ is
$0.5\%$, and the
${\rm d}E/{\rm d}x$
resolution is $6\%$ for electrons
from Bhabha scattering. The EMC measures photon energies with a
resolution of $2.5\%$ ($5\%$) at $1$~GeV in the barrel (end cap)
region. The time resolution in the TOF barrel region is 68~ps, while
that in the end cap region is 110~ps. The end cap TOF
system was upgraded in 2015 using multigap resistive plate chamber
technology, providing a time resolution of
60~ps~\cite{etof1,etof2,etof3}; about 83\% of the date used here
benefits from this upgrade. Luminosity~\cite{lum} at each energy point, with a
total uncertainty of about 1.0\%, is given in Table~\ref{tab:mbc}.

Simulated data samples produced with a {\sc
geant4}-based~\cite{geant4} Monte Carlo (MC) package, which
includes the geometric description of the BESIII detector~\cite{detvis} and the
detector response, are used to determine detection efficiencies
and estimate backgrounds. The simulation models the beam
energy spread and initial state radiation (ISR) in the $e^+e^-$
annihilations with the generator {\sc
kkmc}~\cite{ref:kkmc}.
Inclusive MC
samples 40 times the size of data are used to simulate the background contributions. These samples, which contain no signal $D^+_s \to K_1(1270)^0 e^+\nu_e$ and  $D^+_s \to b_1(1235)^0 e^+\nu_e$ decays,
include the production of open charm
processes, the ISR production of vector charmonium(-like) states,
and the continuum processes incorporated in {\sc
kkmc}.
The known decay modes
are modeled with {\sc
evtgen}~\cite{ref:evtgen1,ref:evtgen2} using branching fractions
either taken from the
Particle Data Group~\cite{pdg2020}, and the remaining unknown decays from the
charmonium states with {\sc lundcharm}~\cite{ref:lundcharm1,ref:lundcharm2}.
Final state radiation~(FSR)
from charged final state particles is incorporated using the {\sc
photos} package~\cite{photos}.
The signal decays $D^+_s \to K_1(1270)^0 e^+\nu_e$ and $D^+_s \to b_1(1235)^0 e^+\nu_e$
are simulated using the ISGW2 model~\cite{isgw2}. The $K_1(1270)^0$ is allowed to decay into all subdecays with the $K^-\pi^+\pi^0$ final state and the branching fractions of $K_1(1270)^0$ subdecays quoted from PDG~\cite{pdg2020}.
The $b_1(1235)^{0}$ decays into the $\omega\pi^0$ final state with $\omega\to\pi^+\pi^-\pi^0$.
A relativistic Breit-Wigner function is used to parameterize the resonances $K_1(1270)^0$ and $b_1(1235)^{0}$,
whose masses and widths are fixed to the individual world-average values~\cite{pdg2020}.

\section{ANALYSIS METHOD}

In the $e^+e^-$ collision data taken at center-of-mass energies between 4.128 and 4.226~GeV,
the $D_s^\pm$ mesons are produced mainly via the $e^+e^-\to D_s^{*\pm}D_s^\mp\to \gamma(\pi^0)D_s^+ D_s^-$ process.
In the analysis we adopt the double-tag (DT) method pioneered by the MARK III collaboration~\cite{DTmethod1,DTmethod2}.
In single-tag (ST) candidates, the $D_s^-$ meson is fully reconstructed via one of its hadronic decay modes.
In a DT candidate, the transition $\gamma(\pi^0)$ from $D_s^{*+}$ and the signal decay are successfully selected in the presence of a ST $D^-_s$ meson.
The branching fraction of the signal decay is determined by
\begin{equation}
\mathcal B_{\rm sig}=\frac{N_{\rm DT}}{N_{\rm ST}^{\rm tot} \cdot \bar{\epsilon}_{\gamma(\pi^0)\rm sig}},
\label{eq1}
\end{equation}
where $N_{\rm DT}$ and $N_{\rm ST}^{\rm tot}$ are the yields of the DT events and of the ST $D^-_s$ mesons, respectively;
$\bar{\epsilon}_{\gamma(\pi^0) {\rm sig}} {=
\sum_i (N^i_{\rm ST} \epsilon^i_{\rm DT}/\epsilon^i_{\rm ST})/N^{\rm tot}_{\rm ST}}$ is the effective signal
efficiency of selecting the $\gamma(\pi^0)$ and the signal decay in the presence of the ST $D^-_s$ meson,
averaging the values obtained for each $i$-th ST mode according to the corresponding yields of ST $D^-_s$ mesons,
where $\epsilon^i_{\rm ST}$ and $\epsilon^i_{\rm DT}$
are the detection efficiencies of the ST $D_s^-$ mesons and the DT candidates (ST and DT efficiencies) for the $i$-th ST mode, respectively.

\section{ST SELECTION}

To reconstruct {the} candidates for ST $D^-_s$, we use thirteen hadronic decay modes:
$D^-_s\to K^+K^-\pi^-$,
$K^+K^-\pi^-\pi^0$,
$\pi^+\pi^-\pi^-$,
$K_S^0K^-$,
$K_S^0K^-\pi^0$,
$K_S^0K_S^0\pi^-$,
$K_S^0K^+\pi^-\pi^-$,
$K_S^0K^-\pi^+\pi^-$,
$\eta_{\gamma\gamma}\pi^-$,
$\eta_{\pi^+\pi^-\pi^0}\pi^-$,
$\eta'_{\pi^+\pi^-\eta_{\gamma\gamma}}\pi^-$,
$\eta'_{\gamma\rho^0}\pi^-$ and
$\eta_{\gamma\gamma}\rho^-$. Throughout this paper, the $\rho$ denotes $\rho(770)$ and the subscripts of $\eta^{(\prime)}$ denote the reconstructed decay modes.

All the charged tracks, except for those from the $K^0_S$ mesons, are required to originate from a region defined as
$|\!\cos\theta|<0.93$, $|V_{xy}|<1$ cm and $|V_{z}|<10$ cm,
where $\theta$ is the polar angle with respect to the $z$ axis (the MDC symmetry axis)
$|V_{xy}|$ and $|V_{z}|$ are the distances of closest approach with respect to the interaction point (IP), in the transverse plane and
along the $z$ axis, respectively.

Particle identification (PID) of
charged kaons and pions is implemented by combining
${\rm d}E/{\rm d}x$ and TOF information. For charged kaon (pion)
candidates, the likelihood for the kaon (pion) hypothesis
is required to be larger than that for the pion (kaon) hypothesis.

The $K_S^0$ candidates are reconstructed via the $K^0_S\to \pi^+\pi^-$ decay.
The two charged pions are required to satisfy $|V_{z}|<20$ cm and $|\!\cos\theta|<0.93$.
They are assumed to be $\pi^+\pi^-$ without any PID selection, and their invariant mass is required to be within $(0.486, 0.510)$ MeV$/c^2$.
The decay length of the $K_S^0$ candidates from the IP is required to be greater than twice the vertex resolution.

Photon candidates are identified using showers in the EMC. The deposited energy of each shower must be more than 25~MeV in the barrel region ($|\cos \theta|< 0.80$) and more than 50~MeV in the end cap region ($0.86 <|\cos \theta|< 0.92$).
To exclude showers originating from
charged particles,
the angle subtended by the EMC shower and the position of the closest charged track at the EMC
must be greater than 10 degrees as measured from the IP.
To suppress electronic noise and showers unrelated to the event, the difference between the EMC time and the event start time is required to be within
[0, 700]\,ns.

The $\pi^0$ and $\eta_{\gamma\gamma}$ candidates  are reconstructed from %$\gamma\gamma$
{photon} pairs with invariant masses being in the mass intervals $(0.115,\,0.150)$ and $(0.500,\,0.570)$\,GeV$/c^{2}$, respectively.
A 1-constraint kinematic fit is performed on each selected %$\gamma\gamma$
photon pair, by constraining the invariant mass to the $\pi^{0}$ or $\eta$ nominal mass~\cite{pdg2020};
to suppress the combinatorial background, $\chi^2<20$ is required.
To select  $\rho^{-(0)}$, $\eta_{\pi^0\pi^+\pi^-}$, $\eta^\prime_{\eta\pi^+\pi^-}$, and $\eta^\prime_{\gamma\rho^0}$ candidates,
the invariant masses of the {$\pi^-\pi^{0(+)}$}, $\pi^0\pi^+\pi^-$, $\eta\pi^+\pi^-$, and  $\gamma\rho^0$ combinations are required to be
within the mass intervals $(0.570,\,0.970)~\mathrm{GeV}/c^2$, $(0.530,\,0.570)~\mathrm{GeV}/c^2$,  $(0.946,\,0.970)$ GeV/$c^2$ and $(0.940,\,0.976)~\mathrm{GeV}/c^2$, respectively.
In addition, for $\eta^\prime_{\gamma\rho^0}$ the energy of the $\gamma$  is required to be greater than 0.1\,GeV.

The low momentum pions from $D^{*+}$ decays are suppressed by requiring the momentum of any pion not from $K_S^0$, $\eta$, or $\eta^\prime$ decays to be greater than 0.1\,GeV/$c$. In order to reject the peaking background from $D^-_s\to K^0_S\pi^-$ in the selection of $D^-_s\to \pi^+\pi^-\pi^-$, the invariant mass of any $\pi^+\pi^-$ combination is required to be outside the mass window $(0.468, 0.528)$ GeV/$c^2$.

The backgrounds from non-$D_s^{\pm}D^{*\mp}_s$ processes are suppressed by using the beam-constrained mass of the ST $D_s^-$ candidates,
defined as
\begin{equation}
M_{\rm BC}\equiv\sqrt{E^2_{\rm beam}/c^4-|\vec{p}_{\rm ST}|^2/c^2},
\end{equation}
where $E_{\rm beam}$ is the beam energy and
$\vec{p}_{\rm ST}$ is the momentum of the ST $D_s^-$ candidate in the $e^+e^-$ rest frame.
The $M_{\rm BC}$ is required to be within the intervals listed in Table~\ref{tab:mbc}.
This requirement retains most of the $D_s^-$ and $D_s^+$ mesons from $e^+ e^- \to D_s^{*\mp}D_s^{\pm}$.

\begin{table}[htbp]
	\centering\linespread{1.15}
	\caption{The integrated luminosities and requirements of $M_{\rm BC}$ for various energy points.}
	\small
	\label{tab:mbc}
	\begin{tabular}{lcr}
		\hline\hline
		$E_{\rm cm}$ (GeV) &Luminosity ($\rm pb^{-1}$) & $M_{\rm BC}$ (GeV/$c^2$) \\
		\hline
		4.128       & 401.5   & $[2.010,2.061]$         \\
		4.157       & 408.7   & $[2.010,2.070]$         \\
		4.178       & 3189.0  & $[2.010,2.073]$         \\
		4.189       & 569.8   & $[2.010,2.076]$         \\
		4.199       & 526.0   & $[2.010,2.079]$         \\
		4.209       & 571.7   & $[2.010,2.082]$         \\
		4.219       & 568.7   & $[2.010,2.085]$         \\
		4.226       & 1091.7  & $[2.010,2.088]$         \\
		\hline\hline
	\end{tabular}
\end{table}

\begin{figure*}[htbp]
\centering
\includegraphics[width=0.75\textwidth]{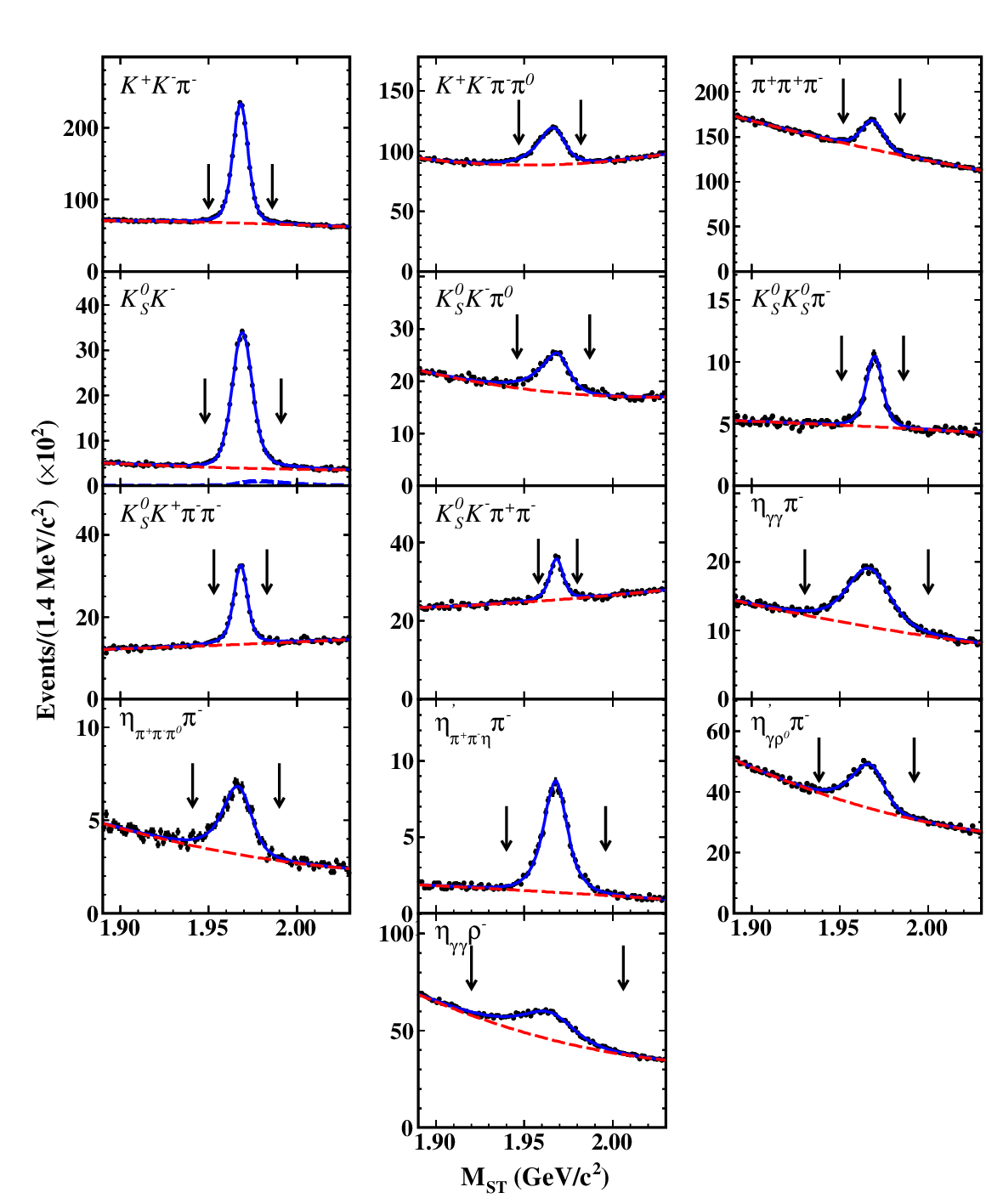}
\caption{\footnotesize
Invariant mass distributions of the ST $D^-_s$ candidates for each ST mode, from the data sample at $E_{\rm cm}$ = 4.178 GeV.
The points with error bars denote the data,
the blue solid curves represent the best fit results, and
the red dashed curves stand for the fitted backgrounds.
For the $D^-_s\to K_S^0K^-$ ST mode, the blue dotted curve is the peaking background from $D^-\to K_S^0\pi^-$.
The arrows indicate the chosen $M_{\rm ST}$ signal regions.
}
\label{fig:stfit}
\end{figure*}

If there are multiple candidates present per ST mode per charge,
only the one with the $D_s^-$ recoil mass
\begin{equation}
M_{\rm rec} \equiv \sqrt{ \left (E_{\rm cm} - \sqrt{|\vec p_{\rm ST}|^{2}c^2+m^2_{D^-_s}c^4} \right )^2/c^4
-|\vec p_{\rm ST}|^2/c^2}
\end{equation}
closest to the $D_s^{*+}$ nominal mass~\cite{pdg2020} is kept for further analysis.
The invariant mass distributions ($M_{\rm ST}$) of the accepted ST candidates for each mode are shown in Fig.~\ref{fig:stfit} for the data sample at $E_{\rm cm}$ = 4.178 GeV.
For each ST mode, the yields of reconstructed $D^-_s$ mesons are derived from fits to the
 individual $M_{\rm ST}$ distributions.
In the fits, the signal is described by the shape obtained by signal MC simulation convolved with a Gaussian function to include the resolution difference between data and simulation.
In the fit of the $D_s^-\to K_S^0K^-$ ST mode,
the shape of the  $D_s^-\to K^0_S\pi^-$ peaking background is modeled by the corresponding MC shape convolved with the same Gaussian function used for the signal shape, and %its
the corresponding yield is left unconstrained. The combinatorial background is described by a second-order polynomial, as verified by analyzing the inclusive MC samples, whose  parameters are left free in the fit.
Figure~\ref{fig:stfit} shows the fit results,
where the black arrows indicate the selected $M_{\rm ST}$ signal regions.
The candidates located in these signal regions are kept for further analysis.

The contribution from the $e^+e^-\to(\gamma_{\rm ISR})D_s^+D_s^-$ process  has been estimated by MC simulation between $0.4\%$ and $1.1\%$, and it has been subtracted to extract the final $D^-_s$ yields.
Table~\ref{tab:ST} summarizes, separately for each ST mode, the total yields of ST $D^-_s$ mesons ($N_{\rm ST}$), obtained summing up all the energy points,
together with the average ST efficiencies
${\epsilon}_{\rm ST} {=
{N_{\rm ST}}/{\sum_j (N^j_{\rm ST}/ \epsilon^j_{\rm ST})}}$,
where $N^j_{\rm ST}$ and $\epsilon^j_{\rm ST}$
are the corresponding ST $D^-_s$ yield and ST efficiency for the $j$-th energy point, respectively.

\begin{table}[htbp]
\centering\linespread{1.15}
\caption{The yields of ST $D^-_s$ mesons ($N_{\rm ST}$) and the average ST efficiencies ($\epsilon_{\rm ST}$) for the complete data sample.
Uncertainties are statistical only.
The efficiencies do not include the branching fractions of the daughter particle decays.}
\small
\label{tab:ST}
\begin{tabular}{l| r@{}l r@{}l }\hline\hline
Tag mode  &\multicolumn{2}{c}{$N_{\rm ST}$~($\times 10^3$)}&\multicolumn{2}{c}{$\epsilon_{\rm ST}$~(\%)}\\ \hline
$K^{+} K^{-}\pi^{-}$                                   &280.7&$\pm$0.9 &40.87&$\pm$0.01  \\
$K^{+} K^{-}\pi^{-}\pi^{0}$                            &86.3&$\pm$1.3  &11.83&$\pm$0.01  \\
$\pi^{-}\pi^{+}\pi^{-}$                                &72.7&$\pm$1.4  &51.86&$\pm$0.03  \\
$K_S^{0} K^{-}$                                        &62.2&$\pm$0.4  &47.37&$\pm$0.03  \\
$K_S^{0} K^{-}\pi^{0}$                                 &23.0&$\pm$0.6  &17.00&$\pm$0.03  \\
$K_S^{0} K_S^{0}\pi^{-}$                               &10.4&$\pm$0.2  &22.51&$\pm$0.05  \\
$K_S^{0} K^{+}\pi^{-}\pi^{-}$                          &29.6&$\pm$0.3  &20.98&$\pm$0.03  \\
$K_S^{0} K^{-}\pi^{+}\pi^{-}$                          &15.3&$\pm$0.4  &18.23&$\pm$0.03  \\
$\eta_{\gamma\gamma}\pi^{-}$                           &39.6&$\pm$0.8  &48.31&$\pm$0.04  \\
$\eta_{\pi^{+}\pi^{-}\pi^{0}}\pi^{-}$                  &11.7&$\pm$0.3  &23.31&$\pm$0.05  \\
$\eta\prime_{\pi^{+}\pi^{-}\eta} \pi^{-}$              &19.7&$\pm$0.2  &25.17&$\pm$0.04  \\
$\eta\prime_{\gamma\rho^{0}} \pi^{-}$                  &50.1&$\pm$1.0  &32.46&$\pm$0.03  \\
$\eta_{\gamma\gamma}\rho^{-}$                          &80.1&$\pm$2.3  &19.92&$\pm$0.01  \\
\hline\hline
\end{tabular}
\end{table}

\section{DT selection}

To reconstruct DT events, the transition photon or $\pi^0$ in the system recoiling against the ST $D_s^-$ is required.
If more than one $\gamma$ or $\pi^0$ candidate is present in the event, only the combination with the smallest energy difference $|\Delta E|$
with $\Delta E \equiv  E_{\rm ST} + E_{\gamma(\pi^0)+D^-_s}^{\rm rec} + E_{\gamma(\pi^0)} - E_{\rm cm}$, is retained, where
$E_{\gamma(\pi^0)+D^-_s}^{\rm rec} \equiv \sqrt{|\vec{p}_{\gamma(\pi^0)}+\vec{p}_{\rm ST}|^{2}c^2 +
  m^2_{D_s^+}c^4}$, $E_i$ and $\vec{p}_i$ ($i = \gamma(\pi^0)$ or ST) are the energy and momentum of
$\gamma(\pi^0)$ or ST $D_s^-$ in the rest initial $e^+ e^-$ frame, respectively.

In the presence of a ST $D_s^-$ and a transition $\gamma\,(\pi^0)$,
the signal decay candidates are selected with the remaining tracks as follows.
There must be exactly three charged tracks available. One of the tracks with charge opposite to that of the ST $D_s^-$ is identified as the positron. The other two oppositely charged tracks are identified as kaon
and pion for $D^+_s \to K_1(1270)^0 e^+\nu_e$, or two pions for $D^+_s \to b_1(1235)^0 e^+\nu_e$.
The selection criteria of kaons and pions are the same as those used in selecting the ST $D_s^-$ candidates.
The {pion} candidate must have charge opposite to that of the positron for $\dstoKev$.
To form a $K_1(1270)^0$ candidate, the $K^+\pi^-\pi^0$ invariant mass is required to be within
$(1.158,1.358)$~GeV/$c^2$. For the $D^+_s\to \pi^+\pi^-\pi^0\pi^0 e^+\nu_e$ candidates,
there are always two possible $\pi^+\pi^-\pi^0$ combinations to form the $\omega$:
the candidate is kept for further analysis if either or both of the combinations has an invariant mass falling in the $\omega$ mass signal region of $(0.757,0.807)$~GeV/$c^2$.
To form a $b_1(1235)^0$ candidate, the $\omega\pi^0$ invariant mass is required to be within
$(1.080,1.380)$~GeV/$c^2$.

The $e^+$ PID uses ${\rm d}E/{\rm d}x$, TOF, and EMC information to construct likelihoods for the electron, pion, and
kaon hypotheses ($\mathcal{L}_e$, $\mathcal{L}_\pi$, and $\mathcal{L}_K$). The $e^+$ candidate is required to satisfy $\mathcal{L}_e>0.001$ and $\mathcal{L}_e/(\mathcal{L}_e+\mathcal{L}_\pi+\mathcal{L}_K)>0.8$. Its deposited energy in the EMC is required to be greater than 0.75 and 0.80 times its momentum reconstructed by the MDC for $D^+_s \to K_1(1270)^0 e^+\nu_e$ and $D^+_s \to b_1(1235)^0 e^+\nu_e$, respectively, to further suppress the background from misidentified hadrons and muons.

The signal candidates are examined by the kinematic
variable
$ {\rm U}_{\rm miss} \equiv |E_{\rm cm} - E_{\rm {ST}} - E_{\gamma(\pi^0)} - E_{h} - E_e| -
|-\vec{p}_{\rm {ST}} - \vec{p}_{\gamma(\pi^0)} - \vec{p}_{h} - \vec{p}_e|c,$ where
$E_k$ and $\vec{p}_k$ ($k = e$ or $h$) are the energy and momentum of the $e^+$ or
the hadron ($K_1(1270)$ or $b_1(1235)$) in the $e^+e^-$ rest frame.
To improve the ${\rm U}_{\rm miss}$ resolution, the candidate tracks, along with the missing
neutrino, are subjected to a
kinematic
fit requiring energy and momentum conservation.
In addition, the invariant mass of each $D_s^{\pm}$ meson
is constrained to the nominal known $D_s^{\pm}$ mass, the
invariant mass of the $D^+_s\gamma (\pi^0)$ or $D^-_s\gamma (\pi^0)$ combination
to the known $D_s^{*\pm}$ mass, and the
combination with the {lowest} $\chi^2$ is kept.
This kinematic fit is only {used} to improve resolution and no event is rejected.
For correctly reconstructed signal events, ${\rm U}_{\rm miss}$ peaks at zero.

To suppress backgrounds from $D^+_s$ hadronic decays, the maximum energy of the
unused showers, $E_{\rm \gamma~extra }^{\rm max}$, must be less than 0.3~GeV and 0.2 GeV for $D^+_s \to K_1(1270)^0 e^+\nu_e$ and $D^+_s \to b_1(1235)^0 e^+\nu_e$, respectively, optimized with the Punzi method~\cite{punzi}
using the inclusive MC samples. The DT candidate events with
additional charged tracks ($N_{\rm extra}^{\rm char}$)
or additional $\pi^0$ candidates, $N_{\rm extra}^{\pi^0} \ne 0$, are removed.

\begin{figure*}[htbp]
\includegraphics[width=0.48\linewidth]{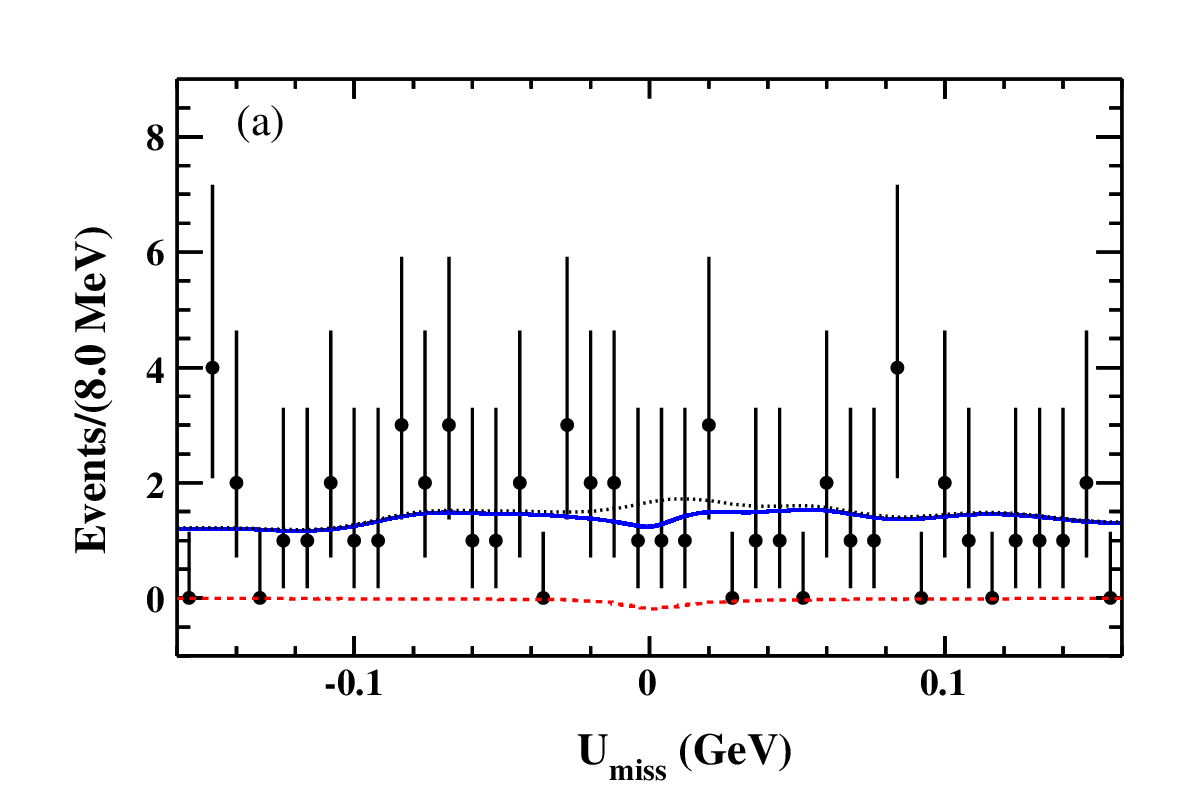}
\includegraphics[width=0.48\linewidth]{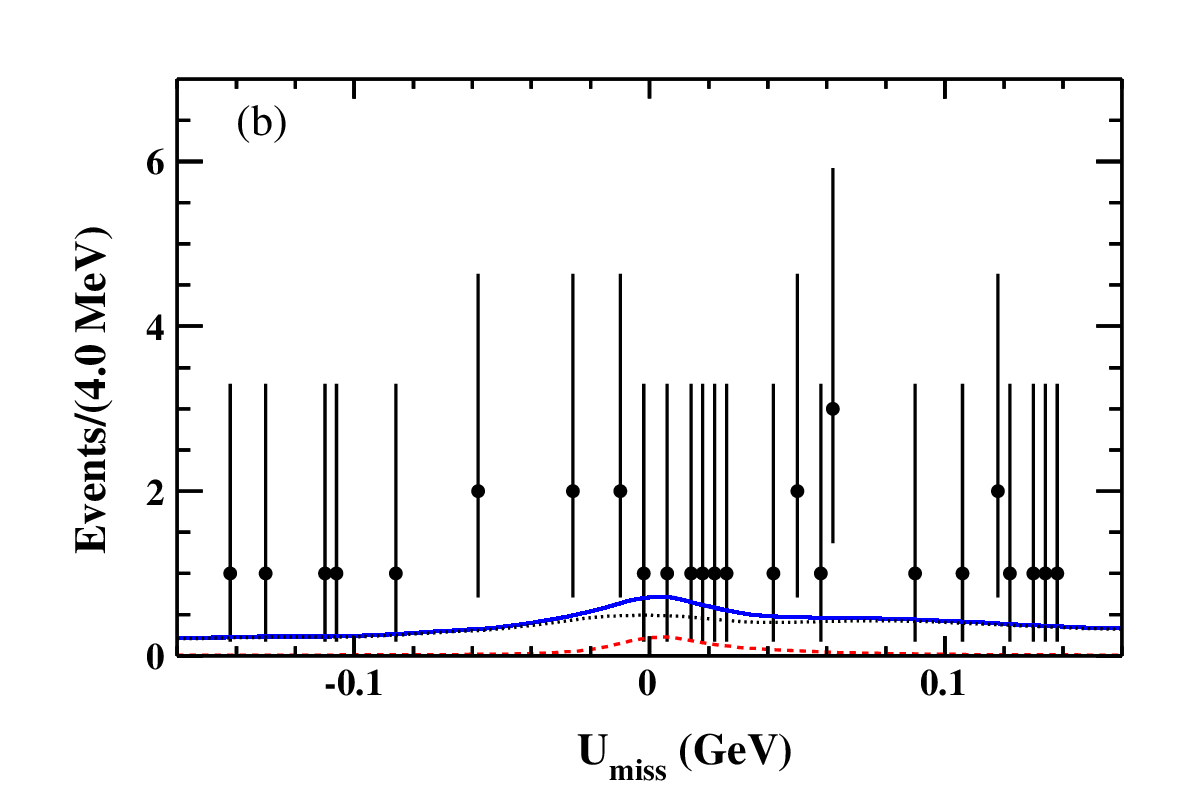}
\caption{Fits to the ${\rm U}_{\rm miss}$ distributions of the
(a) $D^+_s \to K_1(1270)^0 e^+\nu_e$ and (b)  $D^+_s \to b_1(1235)^0 e^+\nu_e$ candidate events.
The points with error bars are data, the red dashed curve is the signal, the black dotted curve is the background contribution, and the blue solid curve shows the total fit.}
\label{fig:fit_Umistry1}
\end{figure*}

Figure~\ref{fig:fit_Umistry1} shows the ${\rm U}_{\rm miss}$ distributions of the selected candidate events in data.
Unbinned maximum likelihood fits are performed on these distributions. In the fits, the signal and background are modeled by the simulated shapes obtained from the signal MC events and the inclusive MC samples, respectively, and the yields of the signal and background are left free.
Since no significant signal is observed, upper limits will be set by assuming all the fitted signals are from $K_1(1270)^0$ or $b_1(1235)^0$.

\begin{table}[htbp]
\centering\linespread{1.15}
\caption{
The DT efficiencies ($\epsilon_{\rm DT}$) and the effective signal
efficiencies ($\epsilon_{\gamma(\pi^0)\rm sig}$) for various ST modes. The uncertainties are statistical only.
All the efficiencies are given in \% and do not include the branching fractions of the daughter particle decays. }
\small
        \label{tab:bf}
        \begin{tabular}{l| r@{}l r@{}l | r@{}l r@{}l }\hline\hline
Tag mode  & \multicolumn{2}{|c}{$\epsilon_{{\rm DT},K_1}$} &\multicolumn{2}{c}{$\epsilon_{{\gamma(\pi^0)\rm sig},K_1}$} & \multicolumn{2}{|c}{$\epsilon_{{\rm DT},b_1}$} &\multicolumn{2}{c}{$\epsilon_{{\gamma(\pi^0)\rm sig},b_1}$}\\ \hline
$K^{+} K^{-}\pi^{-}$                                  &3.10&$\pm$0.03 &7.58&$\pm$0.08 &1.19&$\pm$0.01 &2.92&$\pm$0.03\\
$K^{+} K^{-}\pi^{-}\pi^{0}$                           &0.78&$\pm$0.03  &6.57&$\pm$0.24  &0.28&$\pm$0.01  &2.34&$\pm$0.07 \\
$\pi^{-}\pi^{+}\pi^{-}$                                 &4.21&$\pm$0.04 &8.13&$\pm$0.09 &1.67&$\pm$0.02 &3.22&$\pm$0.04\\
$K_S^{0} K^{-}$                                       &3.78&$\pm$0.05 &7.98&$\pm$0.11 &1.54&$\pm$0.02 &3.25&$\pm$0.05\\
$K_S^{0} K^{-}\pi^{0}$                                  &1.31&$\pm$0.04  &7.74&$\pm$0.22  &0.47&$\pm$0.02  &2.78&$\pm$0.12 \\
$K_S^{0} K_S^{0}\pi^{-}$                               &1.62&$\pm$0.06  &7.19&$\pm$0.26  &0.58&$\pm$0.03  &2.56&$\pm$0.13 \\
$K_S^{0} K^{+}\pi^{-}\pi^{-}$                          &1.33&$\pm$0.05  &6.33&$\pm$0.23  &0.51&$\pm$0.02  &2.43&$\pm$0.09 \\
$K_S^{0} K^{-}\pi^{+}\pi^{-}$                           &1.28&$\pm$0.05  &7.02&$\pm$0.27  &0.42&$\pm$0.02  &2.28&$\pm$0.12 \\
$\eta_{\gamma\gamma}\pi^{-}$                           &4.06&$\pm$0.04 &8.40&$\pm$0.09 &1.67&$\pm$0.03 &3.45&$\pm$0.06\\
$\eta_{\pi^{+}\pi^{-}\pi^{0}}\pi^{-}$                  &1.83&$\pm$0.03  &7.84&$\pm$0.15  &0.70&$\pm$0.02  &2.98&$\pm$0.14 \\
$\eta\prime_{\pi^{+}\pi^{-}\eta} \pi^{-}$              &1.86&$\pm$0.03  &7.39&$\pm$0.13  &0.71&$\pm$0.02  &2.83&$\pm$0.10 \\
$\eta\prime_{\gamma\rho^{0}} \pi^{-}$                   &2.63&$\pm$0.04 &8.10&$\pm$0.14 &1.00&$\pm$0.02 &3.07&$\pm$0.06\\
$\eta_{\gamma\gamma}\rho^{-}$                          &1.83&$\pm$0.06  &9.19&$\pm$0.28  &0.66&$\pm$0.01  &3.30&$\pm$0.07 \\
\hline\hline
\end{tabular}
\end{table}

Table~\ref{tab:bf} summarizes the average DT efficiencies and the effective signal efficiencies for various ST modes, respectively. The average DT efficiency for each ST mode are averaged over the yields of ST $D^-_s$ mesons at the different energy points.
The averaged signal efficiencies $\bar \epsilon_{\gamma(\pi^0)\rm sig}$ are
$0.0773\pm0.0005$ and $0.0295\pm0.0002$ for $D^+_s \to K_1(1270)^0 e^+\nu_e$ and  $D^+_s \to b_1(1235)^0 e^+\nu_e$, respectively.

\section{Systematic uncertainty}

In setting the upper limit branching fractions,
the systematic uncertainties are divided into efficiency independent and dependent parts.

\subsection{Efficiency independent systematic uncertainty}

The efficiency independent systematic uncertainty originates from the fit to the ${\rm U}_{\rm miss}$ distribution of the signal $D_s^+$ decay candidates. It affects the signal yield determination and is dominated by the uncertainty from imperfect knowledge of the background shape. The uncertainty associated with the signal shape is negligible.
This systematic uncertainty
is studied by altering the nominal MC background shape
with two methods. First, alternative MC samples are
used to determine the background shape, where the relative fractions of backgrounds from
$q\bar q$, non-$D_s^{*+}D_s^-$ open charm are varied within their uncertainties.
Second, the
background shape is obtained from the inclusive MC samples using a kernel estimation method \cite{Cranmer:2000du} implemented in RooFit~\cite{ROOT}. The smoothing parameter of
RooKeysPdf is varied between 0 and 2 to obtain alternative background shapes.

\subsection{Efficiency dependent systematic uncertainty}

The efficiency dependent systematic uncertainties in the measurement of the branching fractions are listed in Table~\ref{sys}.

The uncertainty associated with the ST yield $N_{\rm ST}$ is estimated to be 0.5\%, from the change of the $N_{\rm ST}^{\rm tot}$ yield
when modifying the angle of MC-truth association for the signal shape and changing the order of the polynomial for the background shape.
The uncertainties associated with the efficiencies of $e^\pm$ tracking (PID), $K^\pm$ tracking (PID), $\pi^\pm$ tracking (PID),
and $\pi^0$ reconstruction are investigated using data and MC
samples of $e^+e^-\to\gamma e^+e^-$ events and DT $D\bar D$ hadronic
events. The systematic uncertainties from the tracking (PID) efficiencies are assigned as 1.0\% per $K^\pm$, $\pi^\pm$, or $e^\pm$.
The $\pi^0$ reconstruction efficiencies include photon finding, the $\pi^0$ mass window, and the 1-constraint kinematic fit, and the corresponding systematic uncertainty is estimated as 2.0\% per $\pi^0$.

The  uncertainty from the selection of the transition $\gamma(\pi^0)$ from $D_s^{*+}$ with the least $|\Delta E|$ method is estimated to be 1.0\% by using the control samples of $D_s^+\to K^+K^-\pi^+$ and $D_s^+\to\eta\pi^+\pi^0$~\cite{munu}.
The uncertainties of the $E_{\rm \gamma~extra }^{\rm max}$, $N_{\rm extra}^{\pi^0}$, and
 $N_{\rm extra}^{\rm char}$ requirements are estimated to be 2.6\% by analyzing DT $D_s^+D_s^-$ events.
The systematic uncertainty associated with the $\omega$ mass window is assigned to be 1.2\% using a control sample of $D^0\to K_S^0 \omega$~\cite{bes3_D_b1ev}.

The systematic uncertainties related to the $K_1(1270)^0$ and $b_1(1235)^0$ mass windows are estimated using alternative signal MC samples, which are produced by varying the mass and width of the $K_1(1270)^0$ and $b_1(1235)^0$ by $\pm1\sigma$. The maximum changes of the signal efficiencies, 1.5\% and 0.9\%, are assigned as the systematic uncertainties for $D^+_s \to K_1(1270)^0 e^+\nu_e$ and $D^+_s \to b_1(1235)^0 e^+\nu_e$, respectively. The uncertainties due to MC statistics, propagated from those of the ST and DT efficiencies, are 0.8\% and 0.7\% for $D^+_s \to K_1(1270)^0 e^+\nu_e$ and $D^+_s\to b_1(1235)^0 e^+\nu_e$, respectively.

The uncertainty from the quoted branching fraction of the $K_1(1270)^0\to K^+\pi^-\pi^0$ decay is estimated as 10.7\%~\cite{pdg2020,k10}; for the $b_1(1235)^0$ decay the $\omega\to\pi^+\pi^-\pi^0$ uncertainty of 0.8\%~\cite{pdg2020} is considered for the analysis.

The uncertainties due to the signal model are
estimated to be 1.1\% and 1.0\% for $D_s^+\to K_{1}(1270)^{0} e^+\nu_{e}$ and $D_s^+\to b_{1}(1235)^{0} e^+\nu_{e}$, respectively, by comparing the DT efficiencies
obtained from the nominal and phase space models, and varying the relative fractions of different $K_1$ subdecays.

\begin{table}[htp]
\centering\linespread{1.15}
\caption{Efficiency dependent systematic uncertainties of (a) $\dstoKev$ and (b) $\dstobev$. All the
uncertainties are given in \%.
}

      \begin{tabular}{l|c|c}
    \hline\hline
    {Sources}& ~(a)~ & ~(b)~ \\
    \hline
  $N_{\rm ST}$  &   0.5 &   0.5  \\
 $K^{\pm}$, $\pi^{\pm}$ tracking      &2.0 &  2.0        \\
 $K^{\pm}$, $\pi^{\pm}$ PID              &2.0 &   2.0       \\
  $e^{\pm}$ tracking     &1.0 &   1.0       \\
  $e^{\pm}$ PID          &1.0 &   1.0       \\
  $\pi^0$ reconstruction  &2.0& 4.0 \\
  $\gamma(\pi^0)$ from $D_{s}^{*+}$ &1.0 & 1.0 \\
  $E_{\rm \gamma~extra }^{\rm max}$, $N_{\rm extra}^{\rm char}$ and $N_{\rm extra}^{\pi^0}$        & 2.6 & 2.6 \\
  $\omega$ mass window            &-  &  1.2 \\
  $K_1^0(1270)$, $b_1^0(1235)$ mass windows              &1.5  & 0.9  \\
  MC statistics                       &0.8  & 0.7  \\
  $\mathcal B$ of $K_1$ or $\omega$ decays                   &10.7&0.8   \\
  {Signal model}    &1.1  &1.0      \\
        \hline
  Total                                            &11.8&6.2\\
          \hline\hline

            \end{tabular}
\label{sys}
\end{table}

By adding these uncertainties in quadrature, the total efficiency dependent systematic uncertainties, $\sigma_{\epsilon}$, obtained are 11.8\% and 6.2\% for $D^+_s \to K_1(1270)^0 e^+\nu_e$ and  $D^+_s \to b_1(1235)^0 e^+\nu_e$, respectively.

\section{Results}

\begin{figure*}[htbp]
\includegraphics[width=0.48\linewidth]{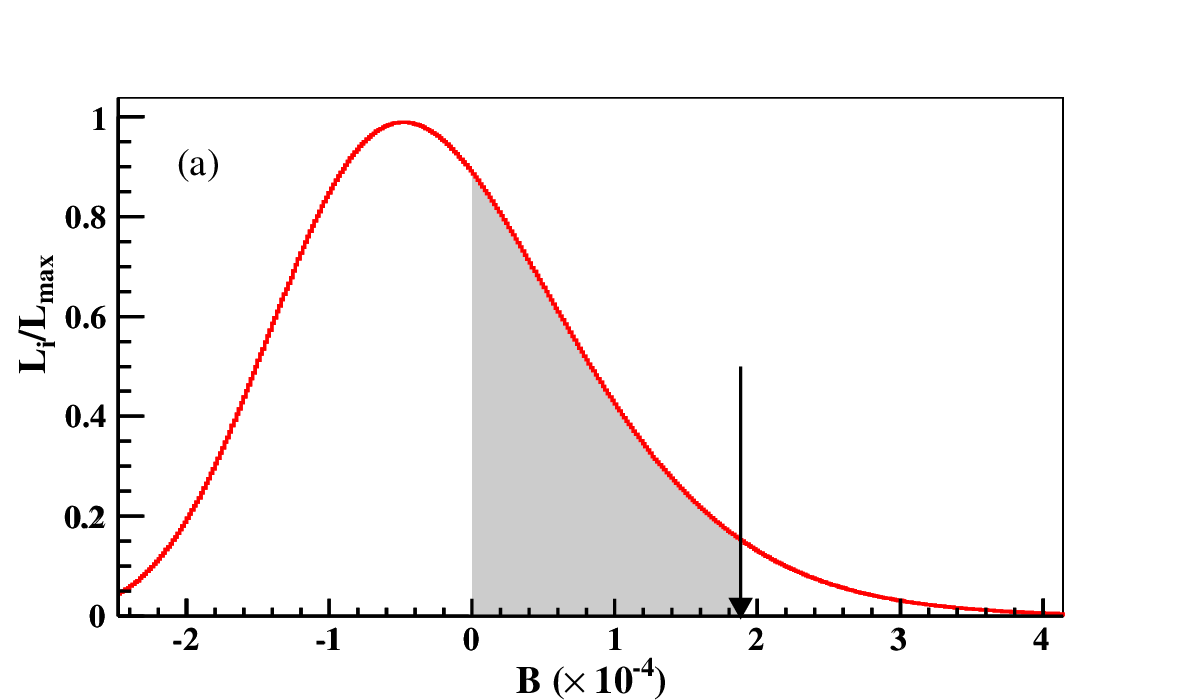}
\includegraphics[width=0.48\linewidth]{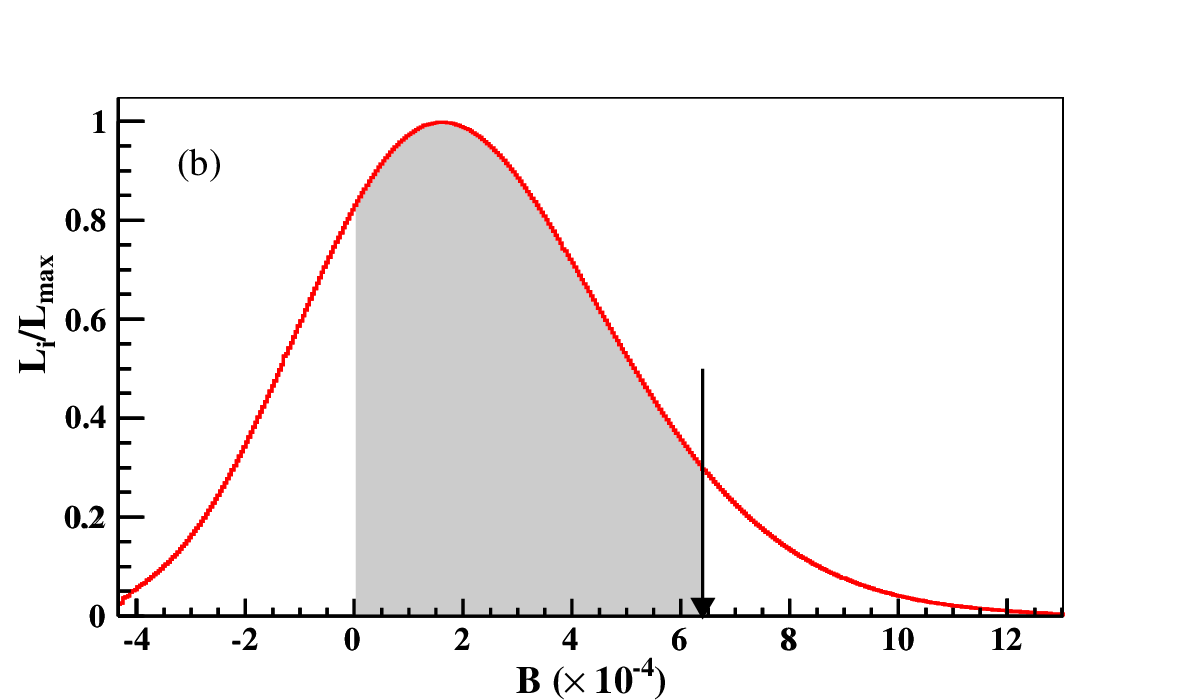}
\caption{The distributions of likelihood versus the corresponding branching fraction products of
(a) $D^+_s \to K_1(1270)^0 e^+\nu_e$ and (b) $D^+_s \to b_1(1235)^0 e^+\nu_e$.
The black arrows correspond to the upper limits at the 90\% confidence level.}
\label{fig:prob}
\end{figure*}

To take into account the efficiency independent systematic uncertainty, the maximum-likelihood fits are repeated using
different alternative background shapes as mentioned in
the previous section and the one resulting in the most
conservative upper limit is chosen. Finally, the second kind of systematic uncertainty $\sigma_{\epsilon}$ is incorporated in the
calculation of the upper limit via~\cite{K.Stenson:2006,cpc:up}
\begin{equation}
L(\mathcal B)\propto \int_{0}^{1}L(\mathcal B\frac{\epsilon}{\epsilon_{0}})\exp\left[\frac{-(\epsilon-\epsilon_{0})^2}{2\sigma^2_{\epsilon}}\right]d\epsilon,
\end{equation}
where $L(\mathcal B)$ is the likelihood distribution as a function of
assumed branching fraction; $\epsilon$ is the expected efficiency and $\epsilon_{0}$ is the
averaged MC-estimated efficiency. The likelihood distributions incorporating the systematic
uncertainties are shown in Fig.~\ref{fig:prob}.

The upper limits on the product of branching fractions at 90\% confidence level, obtained
by integrating $L(\mathcal B)$ from zero to 90\% of the total curve, are
\begin{eqnarray}
{\mathcal B}[{D^+_s \to K_1^0 e^+\nu_e}] \cdot {\mathcal B}[{K_1^0\to K^+\pi^-\pi^0}] < 1.9\times 10^{-4} \nonumber
\end{eqnarray}
and
\begin{eqnarray}
{\mathcal B}[{D^+_s \to b_1^0 e^+\nu_e}] \cdot {\mathcal B}[{b_1^0\to \omega\pi^0}] < 6.4\times 10^{-4}, \nonumber
\end{eqnarray}
where the numbers in the particle notations are omitted for brevity.
Considering the branching fraction of ${\mathcal B}[K_1(1270)^0\to K^+\pi^-\pi^0] = 0.467\pm 0.050$~\cite{pdg2020,k10}, we set an upper limit on
\begin{eqnarray}
{\mathcal B}[{D^+_s \to K_1(1270)^0 e^+\nu_e}] < {4.1 \times 10^{-4}} \nonumber
\end{eqnarray}
at 90\% confidence level, which is comparable with the theoretical prediction in Ref.~\cite{cheng}.

\section{Summary}

In summary, by analyzing 7.33\,fb$^{-1}$  of $e^+e^-$ collision data
collected at $E_{\rm cm}$ between 4.128 and 4.226 GeV with the BESIII detector,
we search for the semileptonic decays $D^+_s \to K_1(1270)^0 e^+\nu_e$ and  $D^+_s \to b_1(1235)^0 e^+\nu_e$ for the first time.
No significant signal is observed for both the decays.
The upper limits on the (product) branching fractions are determined to be
${\mathcal B}[D^+_s \to K_1(1270)^0 e^+\nu_e] < 4.1\times 10^{-4}$ and ${\mathcal B}[D^+_s \to b_1(1235)^0 e^+\nu_e]\cdot {\mathcal B}[b_1(1235)^0\to \omega\pi^0] < 6.4\times 10^{-4}$ at 90\% confidence level.

\section{Acknowledgement}

The BESIII Collaboration thanks the staff of BEPCII and the IHEP computing center for their strong support. This work is supported in part by National Key R\&D Program of China under Contracts Nos. 2020YFA0406400, 2020YFA0406300; National Natural Science Foundation of China (NSFC) under Contracts Nos. 12205175, 11635010, 11735014, 11835012, 11935015, 11935016, 11935018, 11961141012, 12022510, 12025502, 12035009, 12035013, 12061131003, 12192260, 12192261, 12192262, 12192263, 12192264, 12192265, 12221005, 12225509, 12235017; the Chinese Academy of Sciences (CAS) Large-Scale Scientific Facility Program; the CAS Center for Excellence in Particle Physics (CCEPP); CAS Key Research Program of Frontier Sciences under Contracts Nos. QYZDJ-SSW-SLH003, QYZDJ-SSW-SLH040; 100 Talents Program of CAS; The Institute of Nuclear and Particle Physics (INPAC) and Shanghai Key Laboratory for Particle Physics and Cosmology; ERC under Contract No. 758462;
The project ZR2019BA036 supported by Shandong Provincial Natural Science Foundation, China;
European Union's Horizon 2020 research and innovation programme under Marie Sklodowska-Curie grant agreement under Contract No. 894790; German Research Foundation DFG under Contracts Nos. 443159800, 455635585, Collaborative Research Center CRC 1044, FOR5327, GRK 2149; Istituto Nazionale di Fisica Nucleare, Italy; Ministry of Development of Turkey under Contract No. DPT2006K-120470; National Research Foundation of Korea under Contract No. NRF-2022R1A2C1092335; National Science and Technology fund of Mongolia; National Science Research and Innovation Fund (NSRF) via the Program Management Unit for Human Resources \& Institutional Development, Research and Innovation of Thailand under Contract No. B16F640076; Polish National Science Centre under Contract No. 2019/35/O/ST2/02907; The Swedish Research Council; U. S. Department of Energy under Contract No. DE-FG02-05ER41374.

\end{document}